\documentclass[12pt,letter]{article}

\usepackage{graphicx, epsfig, color}
\def\eslt{E_T^{\rm miss}}
\def\to{\rightarrow}

\def\bi{\begin{itemize}}
\def\ei{\end{itemize}}

\def\tH{\tilde H}

\def\tg{\tilde g}
\def\tnu{\tilde\nu}

\def\tq{\tilde q}
\def\tw{\widetilde W}
\def\tz{\widetilde Z}

\def\agt{\stackrel{>}{\sim}}
\def\be{\begin{equation}}  
\def\ee{\end{equation}}  
\newcommand\prd[3]{{\it Phys.\ Rev.\ }{\bf D #1} (#2) #3}
\newcommand\prep[3]{{\it Phys.\ Rept.\ }{\bf #1} (#2) #3}
\newcommand\prl[3]{{\it Phys.\ Rev.\ Lett.\ }{\bf #1} (#2) #3}
\newcommand\plb[3]{{\it Phys.\ Lett.\ }{\bf B #1} (#2) #3}
\newcommand\jhep[3]{{\it J. High Energy Phys.\ }{\bf #1} (#2) #3}
\newcommand\app[3]{{\it Astropart.\ Phys.\ }{\bf #1} (#2) #3}

\newcommand\ijmpd[3]{{\it Int.\ J.\ Mod.\ Phys.\ }{\bf D #1} (#2) #3}
\newcommand\npb[3]{{\it Nucl.\ Phys.\ }{\bf B #1} (#2) #3}
\newcommand\epjc[3]{{\it Eur.\ Phys.\ J. }{\bf C #1} (#2) #3}
\newcommand\ptp[3]{{\it Prog.\ Theor.\ Phys.\ }{\bf #1} (#2) #3}
\newcommand{\hepph}[1]{hep-ph/#1}
\newcommand{\astroph}[1]{astro-ph/#1}

\def\sp{}

\begin{document}
\begin{titlepage}
\begin{flushright}
FSU-HEP/060929\\
UH-511-1095-06
\end{flushright}

\vspace{0.5cm}
\begin{center}
{\Large \bf 
Probing SUSY beyond the reach of LEP2 at
the\\[0.4cm] Fermilab Tevatron: low ${\bf |M_3|}$ 
dark matter models 
}\\ 
\vspace{1.2cm} \renewcommand{\thefootnote}{\fnsymbol{footnote}}
{\large Howard Baer $^{1,2}$\footnote[1]{Email: baer@hep.fsu.edu },
Azar Mustafayev $^3$\footnote[2]{Email: amustaf@ku.edu }, Stefano
Profumo $^4$\footnote[3]{Email: profumo@caltech.edu }
Xerxes Tata $^5$\footnote[4]{Email: tata@phys.hawaii.edu }} \\
\vspace{1.2cm} \renewcommand{\thefootnote}{\arabic{footnote}}
{\it 
1. Dept. of Physics,
University of Wisconsin, Madison, WI 53706, USA \\
2. Dept. of Physics,
Florida State University, Tallahassee, FL 32306, USA \\
3. Dept. of Physics and Astronomy,
University of Kansas, Lawrence, KS 66045, USA \\
4. California Institute of Technology, Mail Code 106-38, Pasadena, CA 91125, USA \\
5. Dept. of Physics and Astronomy,
University of Hawaii, Honolulu, HI 96822, USA
}

\end{center}

\vspace{0.5cm}
\begin{abstract}
\noindent In supersymmetric models where the magnitude of the GUT scale
gaugino mass parameter $M_3$ is suppressed relative to $M_1$ and $M_2$,
the lightest neutralino can be a mixed higgsino-bino state with a
thermal relic abundance in agreement with the WMAP central value for
$\Omega_{\rm CDM} h^2$ and consistent with all other phenomenological
constraints.  In these models, the gluino can be as light as 200 GeV
without conflicting with the LEP2 bounds on the chargino mass. Thus,
gluino pair production can be accessible at the Fermilab Tevatron at
high rates.  In this framework, gluinos decay radiatively with a large
branching fraction to a gluon plus a neutralino. We find that
experiments at the Fermilab Tevatron, with 5 fb$^{-1}$ of integrated
luminosity, will be sensitive to $\tg\tg$ production in the $m_{\tg}\sim
200-350$ GeV range via the multi-jet $+\eslt$ and multi-jet
$+\ell^+\ell^- +\eslt$ channels at the $5\sigma$ level, while trilepton
signatures are expected to be below this level of
detectability. Dilepton mass edges from both $\tz_2$ and $\tz_3$ decays
may be measurable in the dilepton $+$ multi-jet $+\eslt$ channel.

\vspace{0.8cm}

\end{abstract}


\end{titlepage}

\section{Introduction and motivation}

The determination of the average density of cosmological cold dark
matter (CDM)~\cite{wmap}
\begin{equation}
\Omega_{\rm CDM}h^2 = 0.111^{+0.011}_{-0.015} \ \ (2\sigma)\;,
\label{wmap}
\end{equation}
imposes a stringent constraint on any beyond the Standard Model
framework featuring a weakly interacting massive particle stable on
cosmological time-scales.\footnote{We quote the value obtained by the
WMAP collaboration by combining their data with that from the Sloan
Digital Sky Survey.} In particular, (\ref{wmap}) poses a severe
constraint on $R$-parity conserving supersymmetry (SUSY) models where
the lightest neutralino ($\tz_1$) is the lightest supersymmetric
particle (LSP)~\cite{kamion}.

Although it is possible to  reconcile the value of $\Omega_{\rm
CDM}h^2$ determined by the WMAP team~\cite{wmap} with the thermal
neutralino relic abundance $\Omega_{\tz_1}h^2$ expected in the
framework of the minimal supergravity (mSUGRA) model~\cite{msugra},
agreement with (\ref{wmap}) is obtained only within narrow regions, most
of which are close to the boundary of the allowed parameter space. 
While the smallness of these regions reflects the impressive precision
achieved in the determination of $\Omega_{\rm CDM}$, the fact that they
lie close to phenomenologically constrained portions of the parameter
space reflects a general result in the mSUGRA setup: except in the case
where sparticles are light (the so-called bulk region), $\Omega_{\rm
CDM}h^2$ is considerably smaller than the typical mSUGRA expectation
for $\Omega_{\tz_1}h^2$.  Special neutralino annihilation
mechanisms can, however, be operative in the Early Universe, enhancing
the LSP pair annihilation rate and consequently suppressing its relic
abundance to acceptable values.  In mSUGRA, instances of such
mechanisms are resonant neutralino annihilations through $s$-channel
Higgs exchange diagrams~\cite{Afunnel}, the edges of parameter space
where the LSP co-annihilates~\cite{coann} with either a light
stau~\cite{stau} or a light stop~\cite{stop}, or where $|\mu|$ is small
enough so that the  LSP features a substantial higgsino component
(the hyperbolic branch/focus point (HB/FP) region)~\cite{hb_fp}. Several
groups have examined the signals expected in collider experiments,
as well as via direct and indirect searches for neutralino dark matter
in underground detectors, assuming that the parameters are in one of
these WMAP-allowed regions of the mSUGRA
model. 

Motivated by the fact that the correlations between the WMAP measurement
and expectations in other experiments may be model-dependent, there have
been a number of recent studies that have relaxed the universality
assumption, that is the hallmark of the mSUGRA framework. Allowing for
non-universal Higgs boson mass (NUHM) parameters allows for an extended
region of {\sp parameter space where resonant annihilations occur or/and 
where $|\mu|$ is sufficiently small}~\cite{nuhm}, while non-universality
{\sp in the} $SU(2)$ and $U(1)$ GUT scale gaugino mass parameters allows
agreement between $\Omega_{\tz_1}$ and $\Omega_{\rm CDM}$ either via {\sp an enhanced wino fraction in the LSP}~\cite{winodm}, 
or via bino-wino coannihilation~\cite{bwca}. 
These extended scenarios can be distinguished
from one another, as well as from the minimal mSUGRA scenario, because
they {\sp give rise to different outcomes} for collider signals and {\sp for the anticipated detection rates at} dark matter search experiments. 

Another option to obtain a consistent thermal neutralino relic density
is to {\em reduce} the magnitude of the GUT scale $SU(3)$ gaugino mass
$M_3$ relative to the magnitude of the $SU(2)$ and $U(1)$ gaugino
masses~\cite{belanger,m3dm} (the so-called low $|M_3|$ dark matter model
(LM3DM)).  As explained in Ref.~\cite{m3dm}, a lowered relic density
occurs because {\sp a smaller value for} $|M_3|$ also induces lower
values {\sp for the} squark masses and {\sp the trilinear scalar
couplings} via the {\sp running dictated by the (coupled)}
renormalization group equations (RGEs).  The RGE running also
yields a suppression in the absolute size of the soft breaking Higgs
mass squared parameter $|m_{H_{u}}^2|$, which, in turn, lowers the magnitude of
the weak scale superpotential mass parameter $|\mu |$ (fixed by the
value of $M_Z$), so
that the lightest neutralino develops a significant higgsino component,
{\sp giving rise to} mixed higgsino dark matter (MHDM)\footnote{Although
the gluino mass is reduced from its usual value, the gluino - LSP mass
splitting is still large so that gluino co-annihilation can be safely
neglected in the evaluation of the $\tz_1$ relic
density~\cite{profumo}.}.  Agreement with WMAP is then obtained because
{\sp the} neutralino annihilation {\sp rate is enhanced by larger
annihilation amplitudes into gauge and Higgs boson pairs, and
co-annihilation with the lightest chargino and the next-to-lightest
neutralino further suppress the final LSP relic density}.

In the LM3DM scenario, we {\sp generically} expect {\sp the ratio of the gluino to
lightest chargino} mass $m_{\tg}:m_{\tw_1}$ {\sp to} be {\it smaller} than the
{\sp corresponding value $\sim 3-3.5$ expected} in models with universal GUT
scale gaugino masses and large $|\mu|$.  This ratio is important when
comparing collider searches for sparticles {\sp with LEP and} the Fermilab
Tevatron. Assuming that $m_{\tw_1}-m_{\tz_1}$ is not too small and that
$m_{\tnu} \ge 200$~GeV, consistency with LEP2 experiments requires
$m_{\tw_1}>103.5$ GeV. For models with gaugino mass unification and
large $|\mu |$, this bound implies that gluinos must have mass greater
than $\sim 300-350$ GeV.  Such large mass gluinos are difficult to
search for at the Tevatron, {\sp as} their production cross {\sp section is} rapidly suppressed {\sp with increasing masses}.

In the LM3DM model, {\sp instead, relatively light gluinos} (values {\sp
of the gluino mass} as low as $m_{\tg}\sim 200$ GeV would be consistent
with {\sp the} LEP2 constraints) can be copiously produced in hadronic
collisions, and the currently operating Fermilab Tevatron is the obvious
facility to search for these new matter states. {\sp To-date},
experiments at the Tevatron have searched for gluinos in their multi-jet
plus $\eslt$ data sample, and exclude gluinos lighter than {\sp roughly}
200~GeV, irrespective of the squark mass, from their analysis of the Run
1 data~\cite{gluinolim}. Very recently, the D\O \ collaboration, from an
analysis of 310~fb$^{-1}$ of data has obtained a new limit~\cite{dzero}
of $m_{\tg} > 233$~GeV. Unlike the multi-lepton plus jets plus $\eslt$
analyses based on cascade decays of gluinos, inclusive $\eslt$ analyses
are largely independent of the details of the spectrum in the
electroweak {\sp ``-{\em ino}''} sector\footnote{These analyses are not
completely independent of chargino and heavier neutralino masses because
the transverse momenta of the $\tz_1$ LSPs, and hence the $E_T^{\rm
miss}$ spectrum, does depend on the cascade decay patterns. Moreover,
sometimes a lepton veto is also imposed on the SUSY signal.}.

Within any framework with unification of gaugino masses, $m_{\tg} \sim
(3-3.5) m_{\tw_1}$, and the published limits from CDF and D\O \ 
are pre-empted by the LEP limit $m_{\tw_1}\agt 103$~GeV on the chargino mass. 
Within the LM3DM scenario, {\sp instead}, the gluino is relatively light, and the
impact of the LEP chargino limit on the Tevatron gluino search is
clearly reduced, so that it is possible that data from Tevatron
experiments may probe a range of {\sp the} LM3DM model parameter space not
accessible {\sp to} LEP2, either in the current data sample, or in the data
{\sp sample expected} to be accumulated at the Tevatron before the
LHC completes about a year of operation.

In this study we explore the prospects for detection of gluino pair
production within the framework of the LM3DM scenario. The {\sp
remainder of the} paper is organized as {\sp follows. In Sec. 2 we
discuss the parameter space and sparticle mass spectra expected in the
LM3DM model.  In Sec. 3, we discuss signal rates and backgrounds for
gluino pair discovery in the jets $+\eslt$ channel. In Sec. 4, we show
that discovery in the clean trilepton $+\eslt$ channel is unlikely. In
Sec. 5, we show that detection in the dilepton plus multi-jet $+\eslt$
is {\sp a viable possibility}, and that the associated $m(\ell^+\ell^-
)$ distribution can give the characteristic mass edges indicative of the
$m_{\tz_2}-m_{\tz_1}$, and, possibly, also of the $m_{\tz_3}-m_{\tz_1}$
mass difference.  {\sp Finally,} we summarize our results in Sec. 6.

\section{The low ${\bf |M_3|}$ dark matter model}

The low $|M_3|$ dark matter model differs from mSUGRA only in that the
GUT scale gluino mass parameter $M_3$ {\sp needs} not be equal to
$m_{1/2}=M_1=M_2$. The parameter space of this model is thus given by,
\begin{equation} 
m_0, m_{1/2}, M_3, A_0, \tan\beta, sign(\mu),
\label{eq:par}\end{equation} 
where $m_{1/2}$ is taken to be positive without loss of generality, but
$M_3$ can take either sign. For any {\sp set of values} for the {\sp parameters in} (\ref{eq:par}), we can vary $r_3 \equiv M_3/m_{1/2}$ so as to increase
the higgsino content of the LSP {\sp and to drive the} LSP
annihilation rate {\sp to yield a relic LSP density} $\Omega_{\tz_1}h^2$ in
agreement with (\ref{wmap}). In order to get $|\mu|$ small enough, we
must {\sp``{\em slow down}''} the RG evolution of $m_{H_u}^2$ from
its GUT scale value of $m_0^2$ to a negative value at the weak scale --
remember that $m_{H_u}^2({\rm weak}) \sim -\mu^2$ as long as $\tan\beta$
is not very small -- which, in turn, requires a smaller value of $X_t
\equiv m_{Q_3}^2+m_{t_R}^2+m_{H_u}^2+A_t^2$ than in mSUGRA. Since gauge
coupling effects always {\it increase} squark mass parameters as they
evolve from the GUT scale {\sp down} to the weak scale, and since the large $SU(3)$
gauge coupling contributes dominantly to this increase, smaller values
of $X_t$ are obtained by choosing $|M_3({\rm GUT})|$ to be {\it smaller}
than its mSUGRA value of $m_{1/2}$.

We provide a panorama of the LM3DM scenario in Fig.~\ref{fig:mgl}, where
we show contours of fixed gluino mass in the $m_0-m_{1/2}$ plane with
$\tan\beta=10$, $A_0=0$ and $\mu > 0$, {\sp and} where at each point in
this plane $r_3$ has been chosen to obtain the central value {\sp given
in} (\ref{wmap}) for the LSP relic density.  We use Isajet v7.74 for
sparticle mass calculations~\cite{isajet}.  The grey (red) region is
excluded because either electroweak symmetry in not properly broken or
the LSP becomes charged or colored.  The black (blue) region is excluded
by {\sp the LEP2 negative search results for charginos}.  The wiggles in
the plot curves reflect numerical issues related to the precision with
which we require the neutralino relic abundance to saturate the WMAP
central value for the CDM abundance (\ref{wmap}), and also any numerical
instabilities in the code for the determination of $\mu$ as a function
of $r_3$.  On the extreme left of the plot where the gluino mass
contours dive, the $\tz_1$ is dominantly a bino since (due to light
sleptons) the $r_3$ value there need not deviate severely from $\sim 1$.
As we move to larger values of $m_0$ {\sp at fixed $m_{1/2}$}, much
smaller values of $r_3$ are needed {\sp for the neutralino relic
abundance to match the CDM density} in (\ref{wmap}), and we {\sp step}
into the MHDM region which, as explained above, {\sp also features} a
small value of $m_{\tg}$. Indeed we see that for $m_0 \agt 1$~TeV, the
gluino could be lighter than even 200~GeV in a region of parameter space
unconstrained by the negative results of sparticle searches 
at LEP2.
\begin{figure}[!t]
\begin{center}
\epsfig{file=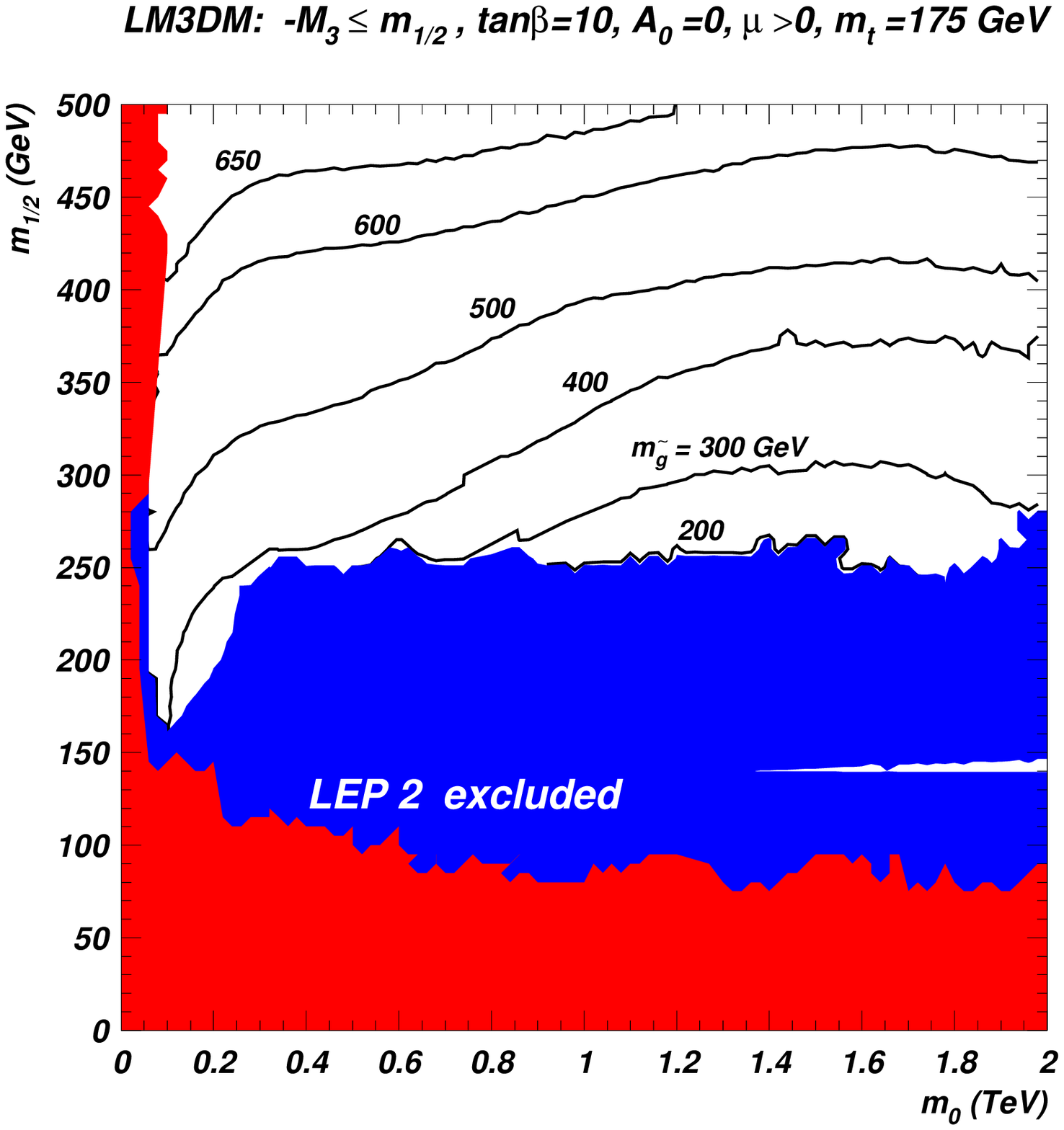,width=8cm}
\epsfig{file=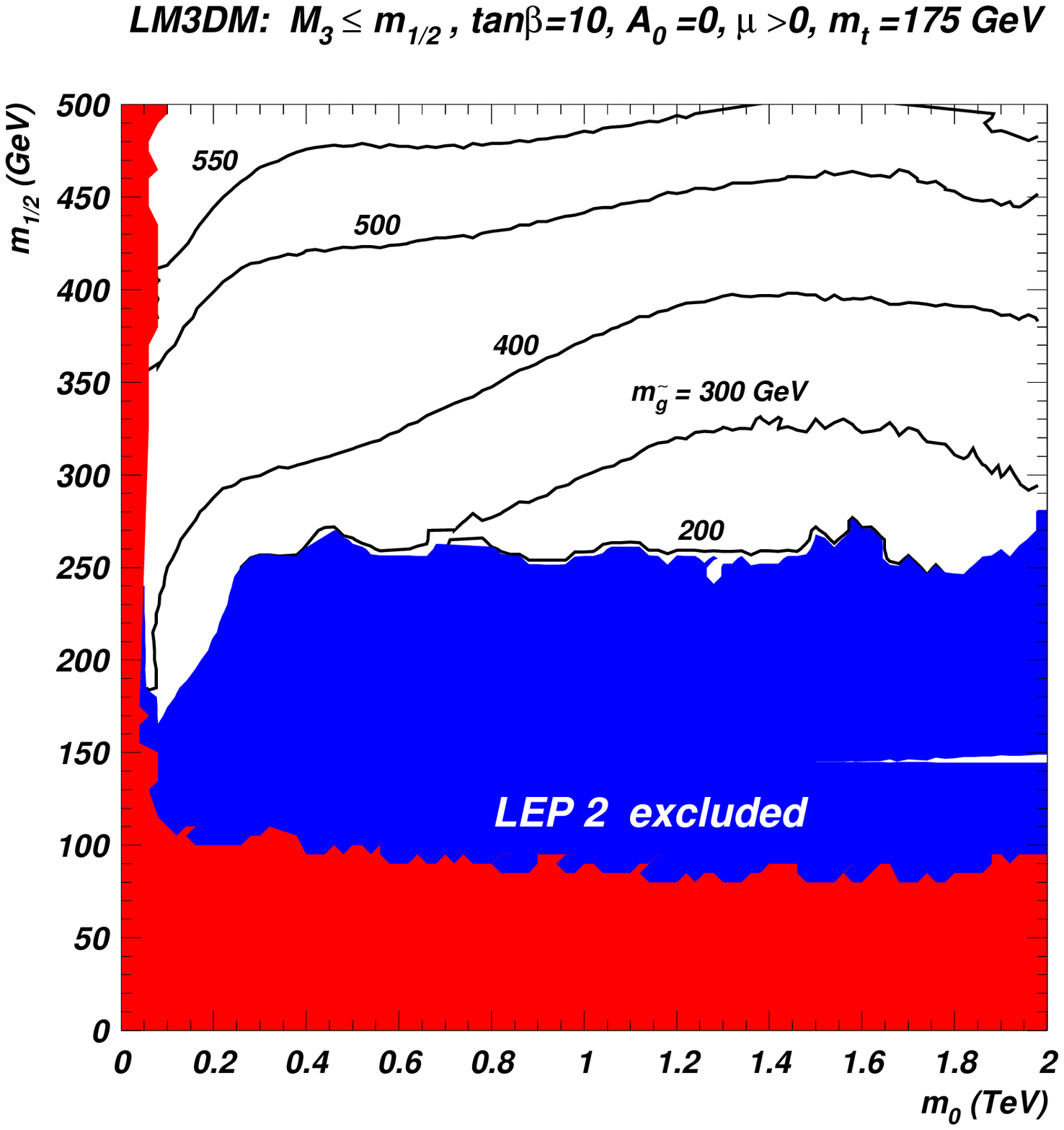,width=8cm}
\end{center}
\caption{\small\it \label{fig:mgl} Contours of $m_{\tg}$ in the $m_0\ vs.\
m_{1/2}$ plane for {\sp the} LM3DM model, where $M_3(M_{GUT})$ has been {\sp set,} at
every point {\sp of the parameter space}, to the value such that $\Omega_{\tz_1}h^2=0.11$.  We take $A_0=0$, $\tan\beta =10$, $m_t=175$ GeV, and {\sp consider $M_3<0$ in the left panel and $M_3>0$ in the right panel}.  The grey (red) regions are excluded because either the electroweak symmetry cannot
{\sp be} correctly broken, or because the LSP is charged.  The black (blue) shaded
regions are excluded by the LEP2 bound on the chargino mass.  }
\end{figure}

As an example of the relation {\sp between} sparticle masses in this
region of parameter space, we show in Fig. \ref{fig:mass} the value of
$m_{\tg}$, together with the chargino and neutralino masses (the
sfermions are too heavy to be accessible at the Tevatron) versus
$m_{1/2}$ for the slice of the plane in Fig.~\ref{fig:mgl} {\sp at
fixed} $m_0=1500$ GeV. This $m_0$ value is representative of the range
needed for which $|M_3({\rm GUT})|$ has to be significantly reduced from
its mSUGRA value in order to obtain agreement with the observed value of
$\Omega_{\rm CDM}h^2$.  While in mSUGRA one expects the masses
$m_{\tg}:m_{\tw_1}:m_{\tz_1}$ to be in the ratio $\sim 7:2:1$, {\sp we
find here that with MHDM, the typical ratio is rather} $\sim 2.5:1.5:1$,
so that not only is the $m_{\tg}-m_{\tw_1}$ mass gap reduced, but the
$m_{\tw_1}-m_{\tz_1}$ mass gap is {\sp suppressed} as well.  Another
noteworthy feature is that because of the {\sp smallness} of $|\mu|$,
there is {\sp sizable} mixing between gauginos and higgsinos resulting
in {\it three} relatively light neutralinos, while the heavy chargino
and the heaviest neutralino (which are dominantly wino-like) are
considerably split from their lighter siblings.  While all the masses
increase steadily with $m_{1/2}$, for the $M_3 > 0$ curves (solid lines)
we see sharp glitches at very low $m_{1/2}\sim 270$ GeV where
$m_{\tz_1}<M_W$: for $m_{1/2} < 270$~GeV, very low values of $r_3$ are
needed since $\tz_1\tz_1\to W^+W^-$ annihilation in the early universe
becomes kinematically suppressed. There are similar glitches for
negative $M_3$ (dashed lines), but these occur for $m_{1/2}$ values
excluded by the LEP2 constraints, and are not seen in the
figure because we terminate the curves on the left when the chargino
mass falls below its LEP2 limit. 
\begin{figure*}[!t]
\begin{center}
\epsfig{file=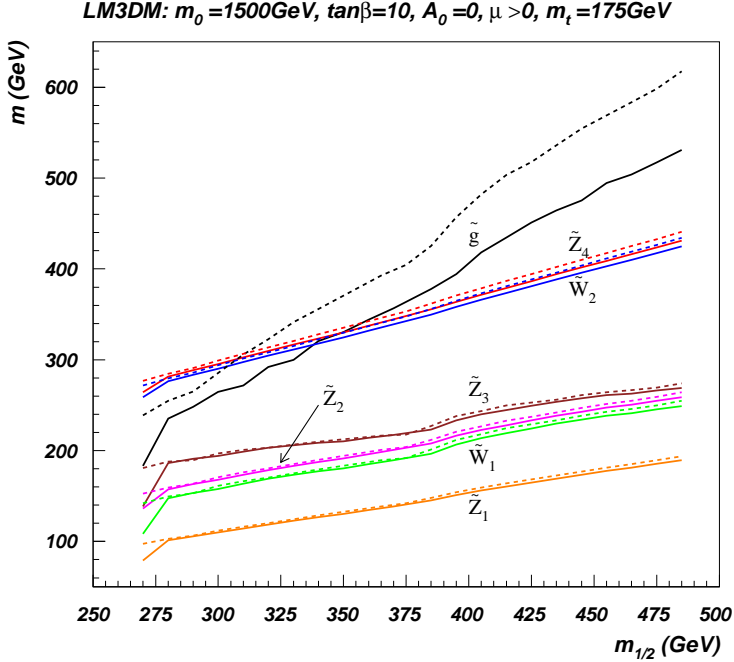,width=10cm} 
\end{center}
\vspace*{-0.4cm}
\caption{\small\it \label{fig:mass} Gluino, chargino and neutralino masses versus
$m_{1/2}$ for the LM3DM model where $M_3(M_{GUT})$ has been lowered at
every point to {\sp obtain} $\Omega_{\tz_1}h^2
=0.11$.  {\sp We take here $A_0=0$, $\tan\beta =10$, $m_t=175$ GeV and $m_0=1500$ GeV}. The solid curves {\sp correspond to} $M_3>0$, while the dashed
curves {\sp to} $M_3<0$. We {\sp cut} the curves on the left when the
chargino mass falls below its LEP2 bound. }
\end{figure*}

{\sp Prior to} discussing whether Tevatron experiments can probe
supersymmetry in this region of parameter space we need to {\sp study}
the decay patterns of the gluino and {\sp of} its daughter
sparticles. For reasons detailed in Ref.~\cite{m3dm}, the radiative
decays $\tg \to g\tz_i$ dominate for the gluino masses of interest at
the Tevatron.  In the upper frames of Fig.~\ref{fig:bfs}, we show the
branching ratio for these various radiative decays of the gluino for
$M_3 >0$ (left frame) and $M_3<0$ (right frame), together with that for
the sum of all its {\sp three-body} decays (labeled 3), versus
$m_{1/2}$. We adopt {\sp here} the same parameter set as in
Fig. \ref{fig:mass}. As {\sp in the preceding figures, we set $M_3$} so
{\sp that $\Omega_{\tz_1}h^2 =0.11$, the WMAP central value for the CDM
abundance~\cite{wmap}}.  We see that -- depending on the sign of $M_3$
-- gluinos lighter than $\sim 420-475$~GeV dominantly decay radiatively.
For small to medium values of $\tan\beta$, where bottom quark Yukawa
couplings can be neglected, the partial width for the various radiative
decays is mainly governed by the $\tH_u$ content of the
neutralino~\cite{m3dm,glrad}, and accounts for the ordering of the
branching fractions for these decays. The sharp rise in the branching
fraction for the three body decays is due to the opening up {\sp of
decays} to the wino-like $\tz_4$ and $\tw_2$, both of which have large
$SU(2)$ gauge couplings to $\tq_L$: when these modes are not phase space
suppressed, they rapidly dominate the decay width. Note that unlike
the tree-level decay, the radiative decay to the dominantly wino-like
$\tz_4$ is dynamically suppressed because the higgsino component of the
wino-like state is always small. Although we have shown the results for
{\sp the particular choice of} $m_0=1.5$~TeV, we have checked that these
results are qualitatively unaltered for $m_0$ values in the range
between 1 and 2~TeV.
\begin{figure*}[!t]
\begin{center}
\epsfig{file=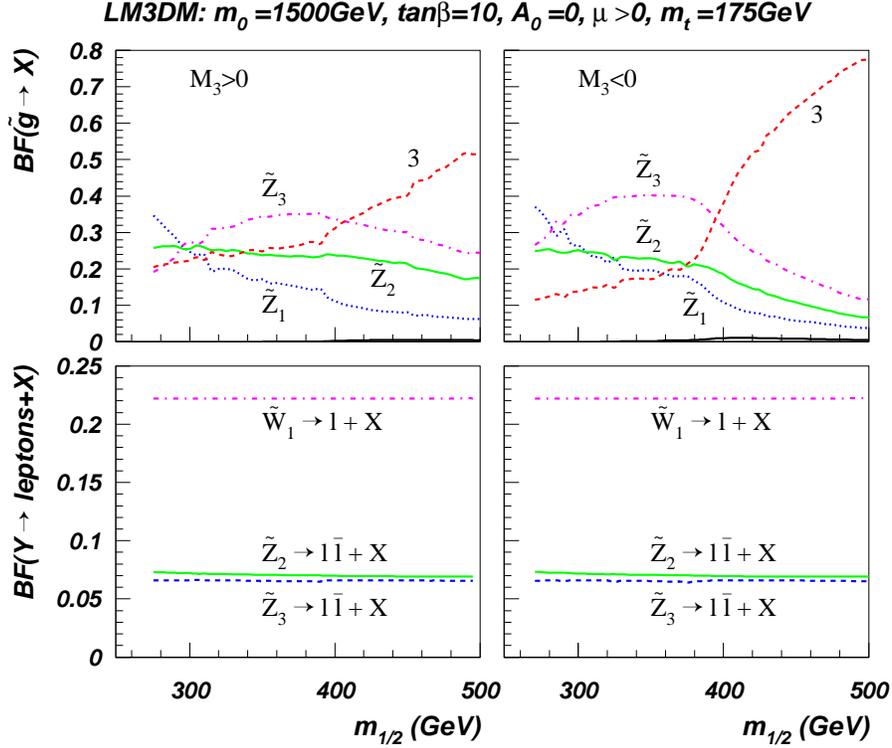,width=12cm} 
\end{center}
\caption{\small\it \label{fig:bfs} The curves (labeled $\tz_i$) in the upper
panels show the branching fractions for the radiative decays $\tg \to
\tz_i g$ of the gluino versus $m_{1/2}$ for LM3DM model where
$M_3(M_{GUT})$ has been adjusted at every point to attain mixed higgsino
dark matter with $\Omega_{\tz_1}h^2 =0.11$, while the curve labeled 3
denotes the corresponding
branching fraction for the sum of all three body decays of
the gluino.  The {\sp (barely visible) unlabeled} lowest curves in {\sp the upper panels indicate} 
$B(\tg \to \tz_4 g)$ which is {\sp find to lie} always below the percent level.
The lower frames show
the total leptonic branching fractions for the decays, $\tz_2 \to
\ell\bar{\ell}+X$, $\tz_3 \to \ell\bar{\ell}+X$ and $\tw_1 \to \ell +X$,
adding in all possible decay chains for the particular lepton topology.
The left (right) frames are for $M_3 > 0$ ($M_3 < 0$).  {\sp Everywhere, we fix}
$m_0=1500$~GeV, $A_0=0$, $\tan\beta =10$ and $m_t=175$.}
\end{figure*}

For large values of $\tan\beta$, the value of $r_3$ required to get
$\Omega_{\tz_1}h^2 =0.11$ is larger; as a result,
the gluino is relatively heavier than $\tz_4$ or $\tw_2$, and its
tree-level decays become dominant. {\sp However, bearing in mind the possibility} that the {\sp cosmological}
dark matter {\sp can very well consist} of several {\sp components, lower values} of
$r_3$ (which would {\sp lead to $\Omega_{\tz_1}<\Omega_{\rm CDM}$}) {\it are not
excluded}. Within the LM3DM framework, Tevatron experiments can, and
should, search for gluinos in this large $\tan\beta$ portion of the
parameter space since it has not been excluded by LEP2 searches. 
By the same token, if dark matter has several components, it is possible
that the {\sp gluinos are even} lighter than what we obtain here (see, {\it e.g.} Fig.~\ref{fig:mass}), {\sp and the resulting LEP2 excluded
region could as well be} smaller than {\sp what we show}.

In the lower frames of Fig.~\ref{fig:bfs} we show the cumulative
leptonic branching fractions for the daughter neutralinos and charginos
versus $m_{1/2}$.  For $\tw_1$ and $\tz_2$, this is simply the usual branching
ratio $B(\tw_1 \to \ell\nu\tz_1)$ and $B(\tz_2\to \ell\bar{\ell}\tz_1)$,
but for $\tz_3$ the two leptons can come from either its primary decay,
or from the leptonic decays of daughter neutralinos\footnote{In
principle, there could be contributions from $\tz_3\to W^\pm \tw_1^\mp$,
but these decays are kinematically inaccessible over the entire {\sp parameter space range shown in the} plot.}.
The branching fractions shown in these lower frames can be used in
conjunction with those in the upper frames
and the gluino production cross sections to
estimate cross sections (before any cuts) for various multi-lepton
topologies in di-jet events at the Tevatron.

\section{Search in the jets ${\bf +\eslt}$ channel}

In this section, we examine whether the Fermilab Tevatron can detect
gluino pair production in the LM3DM model in the multi-jet $+\eslt$ mode,
assuming 5 fb$^{-1}$ of integrated luminosity that is projected to be
accumulated by each experiment at the Tevatron.  We generate signal and background events
using Isajet 7.74, with a toy detector simulation containing hadronic
calorimetry ranging out to $|\eta | <4$, with cell size
$\Delta\eta\times\Delta\phi =0.1\times 0.262$.  We adopt hadronic
smearing of $\Delta E=0.7/\sqrt{E}$ and EM smearing of $\Delta
E=0.15/\sqrt{E}$. We adopt the Isajet GETJET jet finding algorithm,
requiring jets in a cone size of $\Delta R=0.5$ with $E_T^{{\rm jet}}>15$ GeV.
Jets are ordered from highest $E_T$ ($j_1$) to lowest $E_T$.  Leptons
within $|\eta_{\ell}| < 2.5$ ($\ell=e, \ \mu$) are classified as
isolated if $p_T(\ell )>5$ GeV and a cone of $\Delta R=0.4$ about the
lepton direction contains $E_T<2$~GeV. Finally,
if a jet with $|\eta_j|\le 2$ has a $B$-hadron with $E_T \ge 15$~GeV
within $\Delta R \le 0.5$, it is tagged as a $b$-jet with an efficiency
of 50\%. 

To find optimal cuts, we generated 100K signal events for the case where $m_{1/2}=300$ GeV,
$m_0=1500$ GeV, $A_0=0$, $\tan\beta =10$ and $\mu >0$. For this point,
$M_3=79.69$ GeV yields $\Omega_{\tz_1}h^2=0.12$. We have also generated
SM background event samples from $W$ + jets production, $Z$ + jets
production, $t\bar{t}$ production and vector boson pair
production\footnote{We do not estimate QCD backgrounds which, we
assume, are negligible after the cuts described below~\cite{dzero}.}. The
$W$ or $Z$ + jets sample uses QCD matrix elements for the primary parton
emission, while subsequent emissions are generated from the parton
shower.  We adopt a set of cuts similar to those used by the D\O \
collaboration in Ref.~\cite{dzero}, but optimize the $\eslt$ and $H_T$
cut values for this framework. Our final set of cuts are listed in Table
\ref{tab:cuts}, where we divide the signal topologies into $\ge
2$-jets+$\eslt$, $\ge 3$-jets+$\eslt$ and $\ge 4$-jets+$\eslt$, while vetoing
isolated leptons.  The constituent background rates from the major
background sources are listed in Table \ref{tab:bg}. From these rates,
we can compute the signal observability level needed for a given
integrated luminosity, using the following criteria: (1)~the statistical
significance $S/\sqrt{B} \ge 5\sigma$, (2)~$S/B \ge 25$\%, and (3)~$S\ge
10$ events.
%
\begin{table}[!t]
\begin{center}
\begin{tabular}{lccc}
\hline
cut & $2j+\eslt$ & $3j+\eslt$ & $4j+\eslt$ \\
\hline
$\Delta\phi (j_1,j_2)<165^\circ$ & yes & yes & yes \\
isol. lep. veto & yes & yes & yes \\
$n_j$ & $\ge 2$ & $\ge 3$ & $\ge 4$ \\
$|\eta_{j_i}|<0.8$ & $j_1,j_2$ & $j_1,j_2,j_3$ & $j_1,j_2,j_3,j_4$ \\
$80^\circ < \Delta\phi (\eslt ,j_1)<150^\circ$ & yes & yes & yes \\
$\Delta\phi (\eslt ,j_2)$ & $50^\circ - 150^\circ$  & $50^\circ -
150^\circ$ & $60^\circ - 150^\circ$ \\
$\eslt$ & $\ge 120\ {\rm GeV}$ & $\ge 100\ {\rm GeV}$ & $\ge 75\ {\rm GeV}$ \\
$H_T$ & $\ge 220\ {\rm GeV}$ & $\ge 150\ {\rm GeV}$ & --- \\
\hline
\end{tabular}
\end{center}
\caption{\small\it Cuts used for the analysis of multi-jet $+\eslt$ signatures in the 
LM3DM model.
}
\label{tab:cuts}
\end{table}
%

%
\begin{table}[!t]
\begin{center}
\begin{tabular}{lccc}
\hline
BG & $2j+\eslt$ & $3j+\eslt$ & $4j+\eslt$ \\
\hline
$t\bar{t}(175)$ & $6.6\pm 0.3$ & $12.3\pm 0.5$ & $14.9\pm 0.6$ \\
$W+jets$ & $8.9\pm 1.4$ & $15.5\pm 1.9$ & $12.1\pm 1.7$ \\
$Z+jets$ & $11.0\pm 0.7$ & $17.2\pm 0.9$ & $9.0\pm 0.7$ \\
$total$ & $26.5 $ & $45.1 $ & $36.0 $ \\
\hline
\end{tabular}
\end{center}
\caption{\small\it SM backgrounds in ${\rm fb}$ after cuts listed in Table \ref{tab:cuts} for
the multi-jet $+\eslt$ signatures in the LM3DM model.
}
\label{tab:bg}
\end{table}

{\sp Our results} for the SUSY reach of the Tevatron within the LM3DM
framework are shown in Fig. \ref{fig:jets} versus $m_{1/2}$ for the same
parameter choices as in Fig. \ref{fig:mass}. Assuming an integrated
luminosity of 5~fb$^{-1}$, we have checked that the reach in each of the three
$n$-jet + $\eslt$ event topologies is limited by the $5\sigma$
criterion. The minimum cross section for observability of the signal is
shown by the dashed horizontal line, while the signal is indicated by the
solid (dashed) curve for $M_3>0$ ($M_3<0$) for ({\it a})~$\ge
2$-jets+$\eslt$ events, ({\it b})~$\ge 3$-jets+$\eslt$ events and ({\it
c})~$\ge 4$-jets+$\eslt$ events.  We see in each of frames ({\it a})-({\it
c}) that the 5 fb$^{-1}$ reach extends out to $m_{1/2}\sim 330-340$ GeV,
corresponding to a reach in $m_{\tg}$ according to Fig. \ref{fig:mass}
of $\sim 320$ GeV. Within the LM3DM framework, this corresponds to a
reach in $m_{\tw_1}\agt 170$ GeV, and thus extends well beyond that of
LEP2 experiments.
\begin{figure*}[!t]
\begin{center}
\epsfig{file=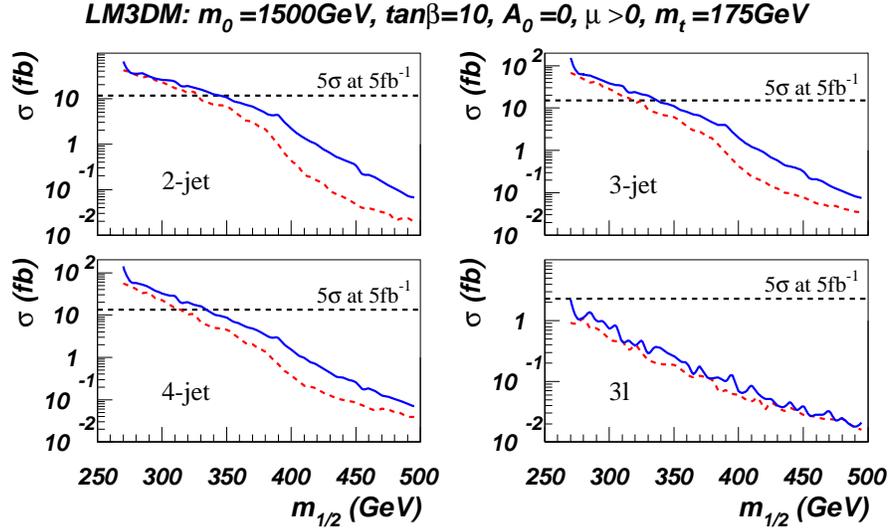,width=12cm} 
\end{center}
\caption{\small\it \label{fig:jets} The SUSY reach of the Fermilab Tevatron with 5
fb$^{-1}$ of integrated luminosity in the multi-jet $+\eslt$ channel for
({\it a}) di-jet events, ({\it b}) tri-jet events, ({\it c} four jet events,
and ({\it d})~trilepton events for LM3DM model where $|M_3(M_{GUT})|$ has
been adjusted at every point to {\sp get} $\Omega_{\tz_1}h^2 =0.11$.  We fix $A_0=0$, $\tan\beta =10$, $m_t=175$
GeV {\sp and $m_0=1500$ GeV}, a slice representative of {\sp the LM3DM parameter space under investigation here}. The solid curve is for $M_3>0$, {\sp while the dashed curve corresponds to} $M_3<0$.  }
\end{figure*}

\section{Search in the trilepton ${\bf +\eslt}$ channel}

We have also examined the reach of the Fermilab Tevatron in the much
touted inclusive trilepton channel\cite{trilep} where the leptons arise from the decays of charginos and
neutralinos produced via $p\bar{p}\to \tw_1\tz_i+X$, or via cascade
decays of gluinos. 
Since, as discussed above, the mass gap between $\tw_1/\tz_2$ and the
$\tz_1$ LSP is not large, we expect the lepton spectra to be relatively
soft.  Hence, for this study, we adopt the soft lepton cuts SC2
introduced in the first paper of Ref.~\cite{new3l}, where the background
was found to be 1.05 fb.  The reach in the inclusive trilepton channel
is shown in Fig.~\ref{fig:jets}({\it d}) where we see that signal is
always below the 5$\sigma$ observability level. This is, in part, due to
the fact that the kinematically favored $\tw_1\tz_{2,3}$ production
now dominantly occurs via the weak isodoublet 
higgsino components of the chargino and
neutralino which have a smaller coupling (than the weak iso-triplet
coupling characteristic of the mSUGRA framework) to the $Z$ boson.  
We conclude that in the
case of the LM3DM model, the best search channel is the multi-jets
$+\eslt$ channels.

\section{Search in the jets + OS-dilepton ${\bf +\eslt}$ channel}

The relatively low value of $|\mu|$ is the characteristic feature of the
LM3DM model. As a result, three (rather than two) neutralinos tend to be
relatively light and mixed, whereas gaugino-higgsino mixing increases the
masses of the 
heavier chargino
and the heaviest neutralino. It is,
therefore, reasonable to ask whether it is possible to identify their
production via the cascade decays of gluinos at the Tevatron. We are
thus led to investigate the observability of the signal in the multi-jet
+ opposite sign (OS) dilepton + $\eslt$ channel, where the leptons have
the same flavor.
This channel is of special importance since it has been long known
that the dilepton invariant mass spectrum from $\tz_i\to
\tz_1+\ell\bar{\ell}$ contains a kinematic cut-off at
$m_{\tz_i}-m_{\tz_1}$. The mass edge(s), if visible, can serve as the
starting point for reconstructing sparticle cascade decays, and for
obtaining information on sparticle masses~\cite{dileptons}.

Toward this end, we examine the signal in the
multi-jet$+\ell\bar{\ell}+\eslt$ channel, where $\ell =e$ or $\mu$.  We
extract signal events containing two opposite-sign/same flavor isolated
leptons plus jets plus missing transverse energy, and compare the signal
with SM backgrounds from $t\bar{t}$ production,
$Z\to\tau\bar{\tau}+{\rm jets}$ production and vector boson pair production
($W^+W^-,\ Z^0Z^0$ and $W^\pm Z^0$ production). By requiring hard
missing $E_T$ ($\eslt >75$ GeV), we reject much of the background from
$Z^0$ production, while by requiring a veto of events with a tagged
$b$-jet we reject much of the $t\bar{t}$ background with hardly any loss
of signal. Finally, requiring at least 2 jets in the events improves the
statistical significance of the signal.  The surviving background rates
in fb, along with signal in the LM3DM framework for $m_{1/2}=300$ GeV
and other parameters as in Fig. \ref{fig:mass}) are listed in Table
\ref{tab:bg2}.  The corresponding reach in the $\ge
2$-jets+$\ell\bar{\ell}+\eslt$ channel is once again governed by the
$5\sigma$ criterion, and is shown in Fig.~\ref{fig:reach_2j2l} versus
$m_{1/2}$, with  other parameters as in
Fig. \ref{fig:mass}, for 5 fb$^{-1}$ of integrated luminosity. We see
that it extends out to $m_{1/2}\sim 310-320$ GeV, {\it i.e.} slightly
lower than the reach in the multi-jet$+\eslt$ channels.
%
\begin{table}[!t]
\begin{center}
\begin{tabular}{lc}
\hline
BG & $2j+\ell\bar{\ell}+\eslt$ \\
\hline
$t\bar{t}(175)$ & $11.6\pm 0.5$ \\
$Z\to\tau\bar{\tau}+jets$ & $5.6\pm 0.5$ \\
$WW,\ WZ,\ ZZ$ & $7.6\pm 0.6$ \\
$total$ & $24.8 $ \\
$signal\ m_{1/2}=300\ GeV$ & $21.4\pm 0.6$ \\
\hline
\end{tabular}
\end{center}
\caption{\small\it SM backgrounds and signal for $m_{1/2}=300$ GeV in ${\rm fb}$ 
after cuts listed in text for
the multi-jet $+\ell\bar{\ell}+\eslt$ signatures in the LM3DM model.
}
\label{tab:bg2}
\end{table}
\begin{figure*}[!t]
\begin{center}
\epsfig{file=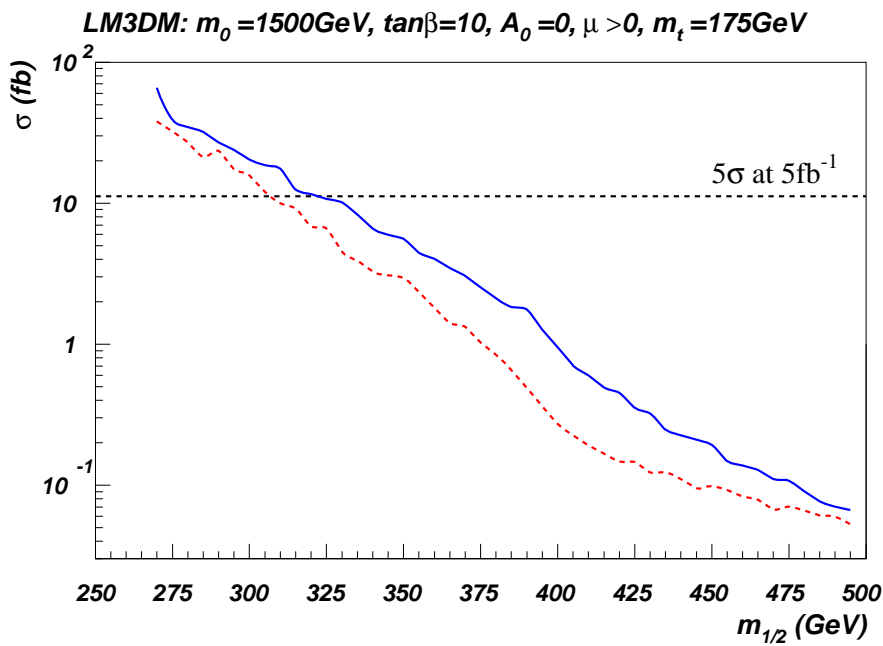,width=12cm} 
\end{center}
\vspace*{-0.8cm}
\caption{\small\it \label{fig:reach_2j2l}
The SUSY reach of the Fermilab Tevatron 
in the $\ge 2~$jets$+\ell\bar{\ell}+\eslt$
channel for the case
of $m_0=1500$ GeV, $A_0=0$, $\tan\beta =10$ and $m_t=175$ GeV assuming
an integrated luminosity of 5~fb$^{-1}$.
We dial $M_3(M_{GUT})$ {\sp for each $m_{1/2}$ so that $\Omega_{\tz_1}h^2
=0.11$}.
}
\end{figure*}

In order to examine the detectability of any dilepton mass edges, we show
the opposite-sign/same-flavor
dilepton invariant mass spectrum from the signal and background
in Fig. \ref{fig:mll}, for
the case of $m_{1/2}=300$ GeV, and other parameters as in Fig.~\ref{fig:mass}.
The hatched distribution comes from the various background sources 
listed in Table \ref{tab:bg2}, which includes a peak at 
$m(\ell\bar{\ell})=M_Z$ from $Z$-pair production. The signal plus 
background is shown by the open histogram. In this case, a
distinct mass edge can be seen at $m_{\tz_2}-m_{\tz_1}\sim 59$ GeV.
Remarkably, the mass edge from $\tz_3\to\tz_1\ell\bar{\ell}$ is also
seen at at $m_{\tz_3}-m_{\tz_1}\sim 86$ GeV.  This higher mass edge will
be somewhat obscured by $Z$-width effects, which are not included in our
simulation of $ZZ$ production.
The point, however, is that the value of
$m_{\tz_3}-m_{\tz_1}$ in our study is only
fortuitously close to $M_Z$, and in general, it may be possible to see
even the second mass edge at the Tevatron! Observation of this second
mass edge would provide a strong {\sp hint for a} small value of
$|\mu|$.

\begin{figure*}[!t]
\begin{center}
\epsfig{file=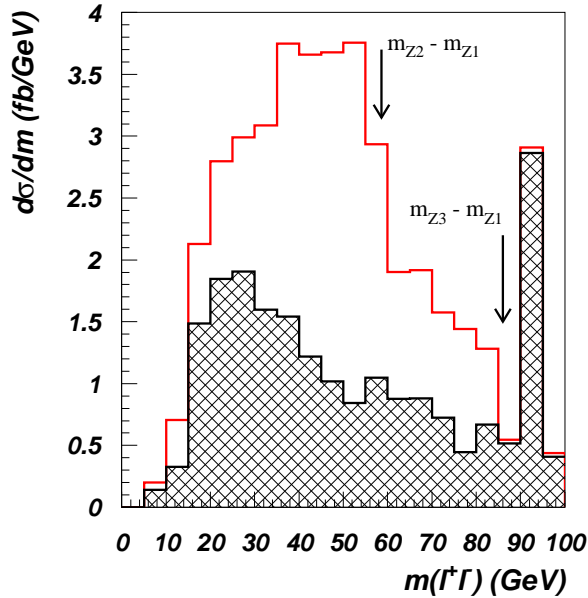,width=8cm} 
\end{center}
\caption{\small\it \label{fig:mll}
The spectrum of opposite sign/same flavor dilepton invariant mass in
background (hatched) and signal-plus-background (open histogram), for the case
of $m_0=1500$ GeV, $m_{1/2}=300$ GeV, $M_3=79.69$ GeV.
We also take $A_0=0$, $\tan\beta =10$ and $m_t=175$ GeV. The arrows
denote the theoretically expected positions of the corresponding mass edges.
}
\end{figure*}

\section{Summary and concluding remarks}

Within the mSUGRA model, or any other supersymmetric setup with 
unification of  the soft SUSY breaking gaugino
mass parameters at the GUT scale, the lower limit $m_{\tw_1} \ge 103$~GeV from
LEP2 experiments excludes gluinos with masses smaller than about
300-350~GeV, leaving little room for gluino searches at the Fermilab
Tevatron. This is, however, a {\it model-dependent} conclusion, and, as
{\sp already} stressed elsewhere~\cite{wss}, Tevatron experiments should
search for gluinos independently of the constraints from LEP2.

{\sp We provide here a specific example,} the so-called low $|M_3|$ dark matter framework
(LM3DM), {\sp where} the universality of the GUT scale gluino mass parameter
with the corresponding $SU(2)$ and hypercharge gaugino mass parameters {\sp is relaxed},
{\sp while the} universality of {\sp all} other soft SUSY breaking parameters {\sp is retained,} as in the
mSUGRA {\sp setup}. Adjusting the magnitude of $M_3({\rm GUT})$ (which can have
either sign) to low values leads to SUSY spectra with relatively
{\sp suppressed values} of $|\mu|$, {\sp entailing, in turn, a larger LSP higgsino fraction}, which {\sp can then lead} to an LSP relic density in
agreement with the {\sp observationally preferred} central value (\ref{wmap}) of $\Omega_{\rm CDM}h^2$ for
any value of {\sp the} other {\sp soft} SUSY breaking parameters. 

The LM3DM framework leads to characteristic differences in the
sparticle spectra from the usually studied frameworks with unified
gaugino masses, or with anomaly-mediated SUSY breaking. In particular,
{\sp low values of $|M_3|$ imply} that {\sp the $m_{\tg}/m_{\tw_1}$ ratio} is
significantly smaller in the LM3DM model {\sp compared to the mSUGRA case,} so that experiments at the
Tevatron will be able to explore regions of parameter space not {\sp already ruled out by} LEP2.

 The main result of {\sp the present} study is the reach of the Fermilab
Tevatron experiments within the LM3DM framework, shown in
Fig.~\ref{fig:jets} and Fig~\ref{fig:reach_2j2l}. {\sp The} best reach
is obtained in the inclusive multi-jet + $\eslt$ channels, while the
reach in the multi-jet plus opposite sign dilepton channel is only
slightly {\sp less effective}. Assuming an integrated luminosity of
5~fb$^{-1}$, {\sp expected} to be {\sp delivered to each experiment}
within the next two years of {\sp operations at the Tevatron}, the reach
extends {\sp up to} $m_{1/2}=350$~GeV which, for $M_3 > 0$ corresponds
to $m_{\tg}\sim 325$~GeV and $m_{\tw_1}\sim 170$~GeV, significantly
beyond the reach of LEP2. Combining the two experiments will yield an
even higher reach.

The concomitant smallness of $|\mu|$ within this framework implies {\sp that
both} $\tz_2$ and $\tz_3$ may be accessible via {\sp gluino decays, offering} another interesting opportunity to Tevatron
experiments, as illustrated in Fig.~\ref{fig:mll}: the invariant dilepton
mass spectrum for events with $\ge 2$ jets + OS dileptons + $\eslt$, with
a veto on $b$-tagged jets (to reduce the background from $t\bar{t}$
production), may yield mass edges from both $\tz_2 \to
\ell\bar{\ell}\tz_1$ and $\tz_3 \to \ell\bar{\ell}\tz_1$
decays. Observation of two mass edges would strongly suggest a small
value of $|\mu|$. 

In summary, if SUSY is realized as in the LM3DM model, a framework
consistent with all constraints from particle physics and cosmology,
experiments at the Tevatron will be able to probe regions of parameter
space not accessible at LEP 2 before the LHC experiments turn on and
collect data for physics analysis. 
We urge our colleagues on the CDF and D\O \ experiments to
search for gluinos irrespective of constraints from chargino searches
since these are based on the {\it untested} 
assumption of gaugino mass unification. 

\section*{Acknowledgments} 
We would like to thank A. Belyaev and J.-F. Grivaz for useful discussions. 
This research was supported in part by grants DE-FG02-97ER41022, 
DE-FG02-95ER40896, DE-FG03-92-ER40701, DE-FG02-ER41291,
and DE-FG02-05ER41361 DE-FG02-04ER41308 from the United
States Department of Energy, NASA NNG05GF69G from NASA 
and the Wisconsin Alumni Research Foundation.
%

\end{document}